\documentstyle[11pt,newpasp,twoside,epsf]{article}
\def\edcomment#1{\iffalse\marginpar{\raggedright\sl#1\/}\else\relax\fi}
\reversemarginpar

\begin{document}
\title{Surface Brightness Fluctuations: a theoretical point of view}

\author{Cantiello$^{1,2}$ M., Raimondo$^{1,3}$ G., Brocato$^1$ E., Capaccioli$^4$ M.}
\affil{$^1$INAF-Osservatorio Astronomico di Collurania - Teramo, Italy} 
\affil{$^2$Dipartimento di Fisica, Universit\`a di Salerno, Salerno, Italy} 
\affil{$^3$Dipartimento di Fisica, Universit\`a di Roma, La Sapienza, Roma, Italy}
\affil{$^4$INAF-Osservatorio Astronomico di Capodimonte, Napoli, Italy}

\begin{abstract}
We present new theoretical evaluations of optical and near-IR
Surface Brightness Fluctuations (SBF) magnitudes for single-burst
stellar populations in the age range $t=5-15$ Gyr and metallicity
from $Z_{\sun}/200$ to $2Z_{\sun}$. Our theoretical predictions
can be successfully used to derive reliable distance evaluations.
They also appear to be a new and valuable tool to trace the
properties of unresolved stellar populations.

\end{abstract}

\section{Introduction}

The SBF method is commonly used to derive accurate extragalactic
distances (Tonry et al. 2001). The basic idea arises from the evidence
that the spatial distribution of the light of nearby galaxies is
``bumpy'', while in more distant ones it appears quite smooth. On CCD
images of galaxies the level of bumpiness was quantified by Tonry \&
Schneider (1988) by defining the apparent SBF magnitudes
$\bar{m}=-2.5~log \bar{f}$, where $\bar{f}$ is the ratio of the pixel
to pixel flux variance to the average pixel flux. This value depends
on the \emph{number} and \emph{kind} of unresolved stars actually
``located'' inside the pixels. Thus, if the absolute SBF magnitudes are
evaluated by using a stellar population synthesis code (as we do), two
major information on distant galaxies can be inferred: \emph{i) the
distance }and \emph{ii) the stellar population properties}.

\section{SBF magnitudes as distance indicator}

On the basis of the stellar population synthesis code developed
at the Teramo Observatory (Brocato et al. 2000), we computed new
absolute SBF magnitudes (UBVRIJHK + HST bands). These predictions are
compared to SBF measurements of objects with SBF-independent distance
estimations. For a metal poor ($Z\leq 0.01$), old ($t=15$ Gyr) population,
our models agree with available data of galactic globular clusters
(Cantiello et al. 2002). To test models of metal rich populations, we
select a sample of galaxies from the Tonry et al. (2001) database of
$\bar{I}$ measurements. Fig. 1 shows the very good agreement between
present models and some observed galaxies with distances derived by
Cepheids or other distance indicators. This supports the reliability
of our models and suggests the use of theoretical SBF magnitudes as
\emph{primary} indicators.

\begin{figure}
\plotfiddle{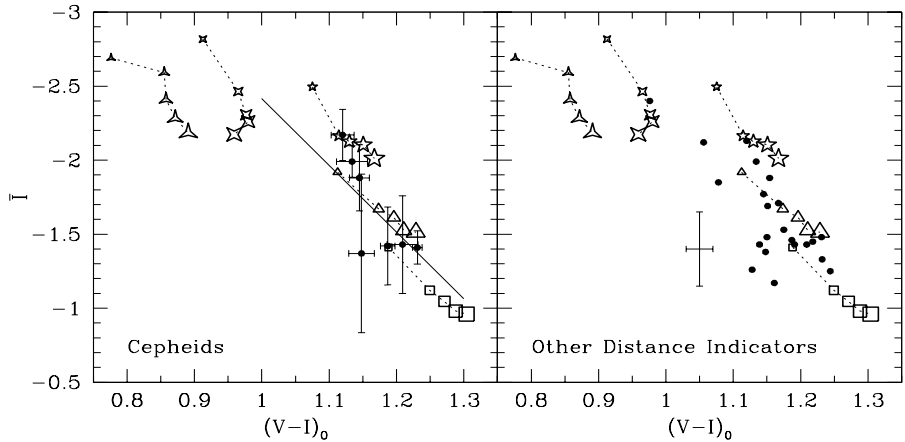}{0in}{0}{60}{55}{-190}{-365}
\vspace{4.6cm} \caption{The comparison between data and new
theoretical SBF models ($Z=0.0001,~0.001,~0.01$: three, four, and
five pointed stars respectively, $Z=0.02$ triangles, $Z=0.04$
squares; symbols with increasing size means increasing age: 5, 9,
11, 13, 15 Gyr). The empirical relation from Tonry et al. is
plotted as a solid line.}
\end{figure}

\section{SBF magnitudes as stellar population tracers}

The SBF signal is very sensitive to the most luminous stars in a
stellar population. As a consequence all the evolutionary phases
have to be properly considered in modeling SBF.  The adopted
population synthesis code is optimized to reproduce the details of
the Color-Magnitude Diagram of star clusters.  Thus, it is
particularly reliable in providing theoretical SBF magnitudes and
colors which can be used to infer the properties (like age and
metallicity) of the stellar population dominating the integrated
light of distant galaxies.

In particular, the $\bar{I}-\bar{K}$ SBF color discloses a
sizeable separation (i.e. larger than typical uncertainties)
between populations with different chemical compositions and ages.
This allows the evaluation of metallicity and age of the studied
stellar population. When used together with the integrated color
$(V-I)$, the resulting two color diagram become a remarkable tool
to investigate unresolved galaxies because it does not depend on
their distances and appears to be  very efficient in removing
the problem of the age-metallicity degeneration (Cantiello et al.
2002).

\end{document}